# Neutron scattering signatures of a quantum spin ice

**Romain Sibille**[1], **Nicolas Gauthier**[2], **Han Yan**[3], **Monica Ciomaga Hatnean**[4], **Jacques Ollivier**[5], **Barry Winn**[6], **Uwe Filges**[2], **Geetha Balakrishnan**[4], **Michel Kenzelmann**[1], **Nic Shannon**[3] **& Tom Fennell**[1]

[1]Laboratory for Neutron Scattering and Imaging, Paul Scherrer Institut, 5232 Villigen PSI, Switzerland, [2]Laboratory for Scientific Developments and Novel Materials, Paul Scherrer Institut, 5232 Villigen PSI, Switzerland, [3]Okinawa Institute of Science and Technology Graduate University, Onna-son, Okinawa 904-0495, Japan, [4]Physics Department, University of Warwick, Coventry, CV4 7AL, UK, [5]Institut Laue-Langevin, CS 20156, F-38042 Grenoble Cedex 9, France, [6]Quantum Condensed Matter Division, Oak Ridge National Laboratory, Oak Ridge, Tennessee, USA, *email: romain.sibille@psi.ch

Quantum spin ice is an appealing proposal of a quantum spin liquid – systems where the magnetic moments of the constituent electron spins evade classical long-range order to form an exotic state that is quantum entangled and coherent over macroscopic length scales. Such phases are at the edge of our current knowledge in condensed matter as they go beyond the established paradigm of symmetry-breaking order and associated excitations. Neutron scattering experiments on the pyrochlore material $Pr_2Hf_2O_7$ reveal signatures of a quantum spin ice state that were predicted by theory.

## INTRODUCTION

Water in its solid form is a peculiar phase of condensed matter: hydrogen atoms occupy the space between tetrahedrally coordinated oxygen atoms, and each oxygen obeys a local rule of forming two long and two short distances with neighbouring hydrogens. This 'ice rule' prevents crystalline water from selecting a single, unique configuration of hydrogen bonds. Instead, water is characterized by a manifold of classical ground states, whose extent just grows exponentially with the sample size. The existence of this manifold of states amounts to a sort of randomness, so that water ice has a 'residual entropy' near zero temperature, which was first realized by Pauling [1] and is of course an interesting fundamental question in the context of the third law of thermodynamics. In spin ices [2], atoms in their lattices are arranged in geometries that resemble that of frozen water, and an analogous local rule for the electronic spins also prevents the formation of a single state of minimal energy – hence the name spin ice.

Discovered in the late 90s in rare-earth pyrochlore oxides, (dipolar) spin ices are materials that contain large magnetic moments distributed on a network of corner-sharing tetrahedra (about 10 $\mu_B$ in the case of $Ho^{3+}$). The structure of the material, via the crystal electric field acting at the rare-earth site, constrains each magnetic moment to align along the local trigonal direction joining its position and the centres of two tetrahedra. The interactions between those large magnetic moments are governed by classical dipole-dipole interactions, which turn out to be effectively ferromagnetic between first neighbours and essentially self-screened on larger pair distances. Such a system – uniaxial magnetic moments that are constrained to their local trigonal direction and interact ferromagnetically on a pyrochlore lattice – has an exact mapping with the problem of hydrogen bonds in water ice. In other words, two of the four magnetic moments in each tetrahedron must be oriented inward and the two others outward. As for water, this '2-in-2-out' local constraint can be achieved in a number of ways that grows exponentially with the number of tetrahedra involved, so that no magnetic order can occur. A spin ice is therefore better viewed as a fluctuating fluid — a spin liquid — of correlated moments, despite its name being inherited from a form of crystalline water ice.

Under certain circumstances, spin liquids – including spin ices – retain dynamic fluctuations between degenerate states even at zero temperature, in which case they are collectively defined as quantum spin liquids [3-4]. This important class of states where the electronic spins lack symmetry-breaking magnetic order has long attracted substantial interest from theorists and experimentalists alike, as they harbour a wealth of exotic physics. Although the initial idea traces back to Anderson's 1973 proposal [5], in which valence bonds between neighbouring spins pair into singlets and resonate on the lattice, the definition of a quantum spin liquid has evolved with years. Such states are now better defined by invoking the long-range entanglement of their ground state wavefunction and the fractionalization of their excitations. In other words, singlets of Anderson's proposal would now form at all pair distances, and excitations can be defined as quasiparticles that cannot be constructed as combinations of the elementary constituents of the system. For instance, the excitations of antiferromagnetic spin-half ($S = 1/2$) chains are deconfined spinons, each carrying $S = 1/2$ – fractionalized quasiparticles that are fundamentally different to the $S = 1$ magnons of conventional three-dimensionally ordered magnets. Neutrons can only excite a pair of spinons, which then propagate on the lattice, so that the existence of such fractionalized excitations leads to a continuum of magnetic excitations in a neutron scattering experiment. A famed example in one dimension is $KCuF_3$ [6]. The physics of fractionalization is also visible in two-dimensional magnets, e.g. kagome-lattice Zn-$Cu_3(OD)_2Cl_2$ [7], honeycomb-lattice α-$RuCl_3$ [8] and triangular-lattice YbMgGaO$_4$ [9]. In each of these materials, the topology plays a fundamental role in the stabilization of an exotic state of matter, as can be directly inferred from their low dimensionalities and lattice names. In three dimensions, precise predictions of quantum spin liquid states exist – such as the quantum spin ice [10] – but experimental realizations remain rather elusive.

## FROM "COULOMB PHASE" TO "MAXWELL PHASE"

Magnetic moments in a spin ice can be seen as local magnetic fluxes B$_i$ that link to form a diamond lattice (Figure 1). The ice rule constrains the sum of the fluxes at each vertex to remain zero. This description based on spin variables can be transformed to a 'continuous medium' by considering the mean value B(r) of the lattice fluxes over a certain volume (volume centred on r, and much bigger than a lattice constant but much smaller than the size of the system). The vector field B(r) obtained from this 'coarse-graining' operation has the physical meaning of a magnetization. Finally, the zero-divergence condition of the local ice rule implies for the vector field that B(r)=∇×A(r), where the vector potential A(r) is an emergent (Coulomb) gauge field. In other words, the magnetic field emerging from the spin ice manifold has the properties of a Coulomb potential.

The description of spin ices in terms of an emergent 'magnetic Coulomb phase' [11] explains, by itself, their originality and beauty among other phases of condensed matter. It also turns out that this approach describes accurately many aspects of their properties. A celebrated example of exotic behaviour in spin-ice materials is that of magnetic monopoles [12]. The magnetic moments in the material interact in such a manner that separate mag-

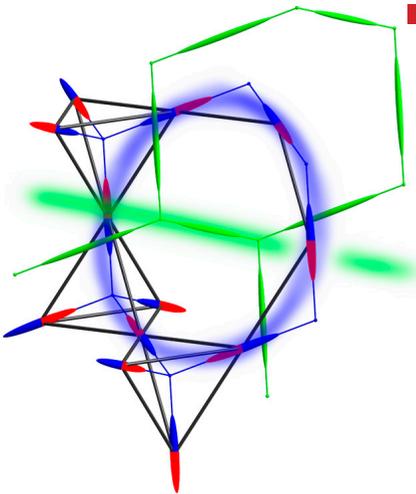

**Figure 1**

In a classical spin ice (CSI), uniaxial magnetic moments decorate a pyrochlore lattice (in black). Magnetic moments (blue/red ellipses) on each tetrahedra are constrained by a local '2-in-2-out' organization principle. Moments can be viewed as magnetic fluxes forming a diamond lattice (in blue), which can be coarse-grained to define a continuous medium with emergent magnetostatics. Quantum dynamics on a six-member ring create electric flux variables (in green) that form a second (interpenetrated) diamond lattice. This quantum spin ice (QSI) ground state can be thought of as a lattice analogue of quantum electrodynamics – making the sample a tiny universe with its own emergent light of gapless magnetic excitations.

netic charges can emerge as 'quasi-particles' associated with excitations. But exotic as these phenomena may be, they can be still fully described with the framework of classical magnetostatics described in the previous paragraph.

An intriguing question is what happens when quantum effects are thrown into the mix [10]. Theoretical works have predicted that quantum-mechanical tunnelling between different spin-ice configurations can lead to excitations that are qualitatively different from those in classical spin ice (CSI) [13]. In the latter case, the quasi-particles associated with the excitations can be thought of as magnetostatic charges. In contrast, in quantum spin ice (QSI), behaviour emerges that is described by quantum electromagnetism. That is, time fluctuations of the gauge field A(r) give rise to an electric field, E(r), so that the emergent field is now a dynamic electromagnetic field. In a more intuitive 'spin language', the dominant tunnelling processes responsible for the time fluctuations of the spin ice manifold are six-member loops – the centre of which define electric flux variables forming a second, interpenetrated diamond lattice (Figure 1). This leads to a rich set of novel phenomena: not only should quantized variants of magnetic monopoles appear in quantum spin ices, but also electric monopoles (equivalent to electric charges) and excitations that behave like photons.

The experimental realisation of a QSI, however, is challenging, and attempts to identify its manifestations have been made in various pyrochlore materials [14]. Nonetheless, spectacular results of neutron scattering experiments hinting at ground states with quantum origins in rare-earth pyrochlores were thereby obtained over the past few years (see [15-16], for example).

# SIGNATURES IN NEUTRON SCATTERING

The free energy of the effective vector field describing a CSI manifold has the form $S \propto \int d^3r\, \mathcal{B}(r)^2$, from which the correlation function can be worked out and written as $\langle S_\mu(-\boldsymbol{k})S_\nu(\boldsymbol{k})\rangle \propto (\delta_{\mu\nu} - \frac{k_\mu k_\nu}{k^2})$. An important consequence for scattering experiments is the existence of special points in k-space, where the limit depends on the direction of approach. These 'pinch-points' are thus singular features directly originating in the two-in-two-out correlations of the ice rule, and are found in a subset of zone centers. Another remarkable point that can be drawn from this correlation function is the dependence in real space of the spin-spin correlations, $C^B_{\mu\nu}(r) \propto \frac{1}{r^3}$, which is not usual for a liquid.

In the case of classical dipolar spin ices such as $Dy_2Ti_2O_7$ and $Ho_2Ti_2O_7$, however, pinch points proved surprisingly difficult to observe. The reason is the scale of the dipolar interactions present. In classical spin ices, correlations are driven by the interplay of single-ion anisotropy and long-range dipolar interactions, and the latter overlay the pinch points with other, highly-structured, scattering which obscures the pinch-point singularity. Ultimately, pinch points were resolved in $Ho_2Ti_2O_7$, through a carefully-tailored application of polarization analysis [17]. These experiments were specifically designed to separate pinch points in the "spin-flip" (SF) channel from the other, non-singular features arising from dipolar interactions in the "non spin-flip" (NSF) channel.

In contrast, in models of CSI driven by nearest-neighbor exchange interactions alone, polarization analysis is not needed to resolve pinch points. And, as Figure 2 shows [18], pinch points can be cleanly resolved in quasi-elastic scattering from $Pr_2Hf_2O_7$ without any need for polarization analysis, which argues strongly against the presence of dipolar interactions. This is expected given the amplitude of the magnetic moments in this compound, about 2.4 $\mu_B$ [19], so that the scale of the dipole-dipole interaction – proportional to the squared moment – reduces to less than 0.1 K, to be compared with about 1.4 K for dipolar spin ices such as $Ho_2Ti_2O_7$. The implication is that the correlated regime appearing below 0.5 K in $Pr_2Hf_2O_7$ [19] must be driven by nearest-neighbour exchange interactions.

We now come to the point of distinguishing between the correlations of a CSI driven by nearest-neighbour exchange interactions and those of an additionally quantum-entangled state such as the QSI model [20-21]. In the latter case, the ground state is governed by the Maxwell action $S \propto \int dt d^3 r \, [\mathcal{E}(r)^2 - c^2 \mathcal{B}(r)^2]$, where c is the speed of light of the emergent photon excitation – a transverse fluctuation of the gauge field. The fact that the spins now fluctuate in time as well as in space introduces an additional power of k in the correlation function, which becomes $\langle S_\mu(-k)S_\nu(k)\rangle \propto k(\delta_{\mu\nu} - \frac{k_\mu k_\nu}{k^2})$. At zero temperature this has the effect of eliminating the pinch points [20]. These, however, were also predicted to be partially restored at finite temperatures in a QSI, and a crossover to a CSI regime is expected at higher temperatures [21]. In $Pr_2Hf_2O_7$, we have carefully analyzed the shape of the pinch points in order to attempt to classify the correlated regime on this temperature scale (Figure 3). It turns out that the line-shape of the pinch point scattering in our experiment is more consistent with $Pr_2Hf_2O_7$ being in the QSI regime at finite temperature than in a nearest-neighbour CSI regime. Another implication of using the field theory of a QSI to model our data, is that the speed of the emergent light can be directly obtained from the fit of the line-shape of the pinch points. It gives a modest 3.6 meters per second, which – besides being the pace of a 4-hour marathon – translates into a bandwidth of 0.01 meV for the corresponding gapless excitations in a neutron scattering experiment. This bandwidth validates a posteriori our analysis, since the quasi-elastic data shown on Figure 2 integrate over ±0.06 meV and the QSI correlations calculated using the field theory contain the photon excitations.

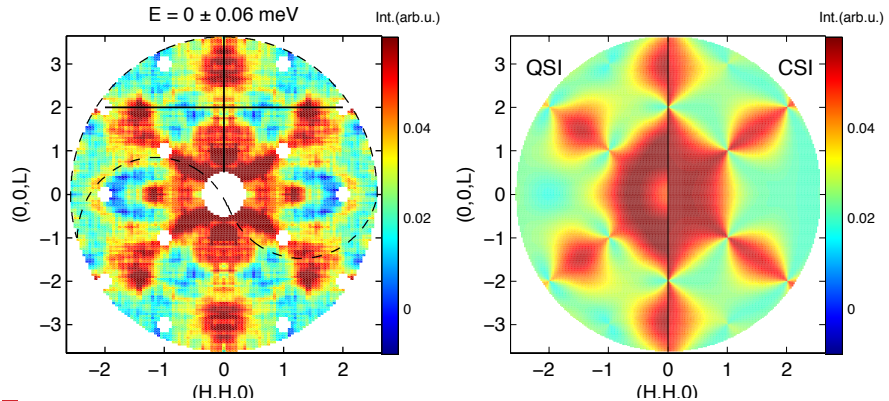

**Figure 2**
The quasi-elastic structure factor of $Pr_2Hf_2O_7$ measured at 0.05 K on IN5 at ILL (left), and the corresponding patterns calculated using a field theory for the classical nearest-neighbour spin ice (CSI) and quantum spin ice (QSI) models [18]. The two calculated patterns are the correlation functions (including the gapless inelastic scattering of the photon excitations) obtained from the best fits to the line-shape of the pinch point scattering around (0,0,2) – as shown in Figure 3.

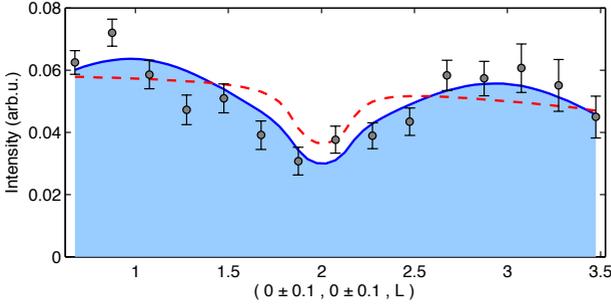

**Figure 3**
Line-shape of the quasi-elastic structure factor of $Pr_2Hf_2O_7$ measured at 0.05 K on IN5 (grey points with error bars), and the corresponding best fits obtained using a field theory for the classical nearest-neighbour spin ice (CSI – red dashed line) and quantum spin ice (QSI – blue line) models [18].

The QSI model is clearly favoured over the CSI model from the fits of the quasi-elastic scattering (Figure 3). However, the statistical difference does not allow to make a definitive conclusion. In order to confirm that the QSI model explains the ground states properties, we have looked at the inelastic scattering at higher energies, which allows to distinguish between QSI and CSI. Indeed, non-dispersing quasiparticle excitations with a unique energy are the only excitation expected from the monopoles in a CSI, while it is well-established that a continuum of scattering is expected from the quantum-coherent quasiparticles of a QSI (gapped magnetic/electric monopoles). The spectrum presented on Figure 4a reveals a broad continuum of excitations present at 0.05 K, whose magnetic origin is assessed by the polarized neutron spectrum (Figure 4b).The latter was recorded on the HYSPEC spectrometer at Spallation Neutron Source (Oak Ridge National Laboratories, USA), using an array of polarization analysers built at the Paul Scherrer Institut (Villigen, Switzerland). The energy spectra taken in the IN5 data confirm that this continuum of spin excitations extends up to at least E = 1 meV – almost an order of magnitude larger than the dominant exchange in the system, which agrees with theoretical predictions [22-24]. At low energy, the spectral weight is peaked around E = 0.2 meV, which is consistent with our data taken on a powder sample at the same temperature [19]. We also notice that the continuum of scattering is qualitatively similar to recent estimates of the spinon contribution to the local dynamical susceptibility [25]. Importantly, the form of scattering found at finite energy in our data (i.e. the "starfish" pattern shown on Figure 4c) differs significantly from the scattering of the quasi-elastic map (Figure 2). The "starfish" scattering at finite energy transfers is highly reminiscent of quantum Monte Carlo simulations for QSI at temperatures where there is a finite density of spinons [26].

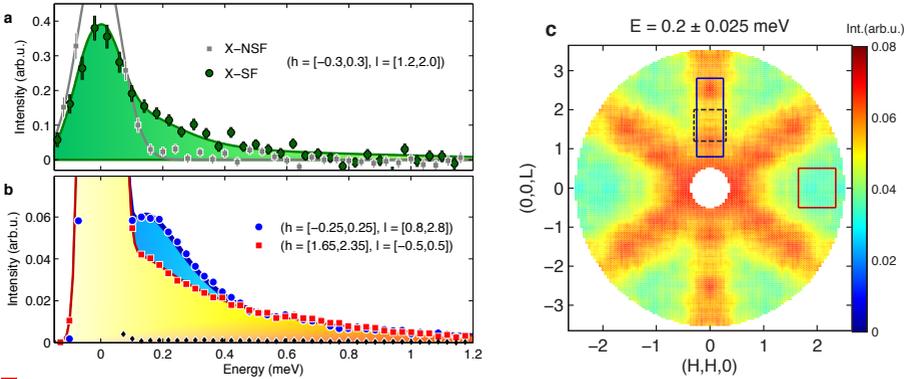

**Figure 4**
Inelastic spectra of $Pr_2Hf_2O_7$ measured at 0.05 K on HYSPEC using polarized neutrons (a) and on IN5 using unpolarized neutrons (b) [18]. The fact that spin flip scattering of neutrons polarized in the horizontal plane (X-SF) is a purely magnetic signal rules out the possibility that these excitations have a non-magnetic origin. The different spectra for the two integration areas in the IN5 data (blue and red rectangles on panel c) reflect the momentum-space dependence of the excitations that show a "starfish" pattern for energy transfers centred around 0.2 meV (panel c).

## MICROSCOPIC ORIGINS

So far, we have essentially described neutron scattering data and compared these with the predictions of a field theory [21], which – by nature – ignores the microscopic details leading to the QSI state. Thus, we now discuss the possible origins of the quantum fluctuations in $Pr_2Hf_2O_7$.

In most rare-earth pyrochlore oxides, at low temperature, the crystal-electric field ground-state doublet is well isolated from the excited levels, so that theoretical models of correlated phases in these materials consider a system of pseudo-spins with $S = ½$ [27-28]. Given the symmetry at the rare-earth site ($D_{3d}$), and depending on the number of f electrons and details of the crystal field, the components of the pseudo-spin can have different properties [14]. In the local trigonal direction of the pseudo-spin (i.e. the direction of the magnetic moments that we defined at the beginning for a classical spin ice), the component of the pseudo-spin always transforms like a magnetic dipole. However, in some cases, components in the other directions transform like electric quadrupoles or magnetic octupoles. It was established by theoreticians that interactions between these higher-rank multipoles can stabilize a QSI [29]. This comes with the condition that multipoles act as a transverse perturbation relative to a dominant ferromagnetic interaction establishing the spin ice manifold. In $Pr_2Hf_2O_7$, we have shown in a previous work that the crystal-electric field around the non-Kramers $Pr^{3+}$ ion promotes a ground-state doublet with a magnetic moment along the trigonal direction, and electric quadrupoles in the plane perpendicular to it [19]. It is therefore likely that transverse exchange interactions between quadrupoles play a role in the correlated ground state of this material.

However, multipolar exchange is not the only way to introduce transverse fluctuations in a material like $Pr_2Hf_2O_7$. In pyrochlores, the ground state of a non-Kramers ion such as $Pr^{3+}$ is doubly-degenerate for reasons of crystal symmetry, but this degeneracy can be lifted by structural distortions. Theoreticians have used this to demonstrate that small amounts of non-magnetic disorder in non-Kramers spin ices – equivalent to a random transverse field – is able to turn a CSI into a QSI ground state [30-31]. Therefore, although samples appear to be of very high quality [32], minute amounts of disorder – as always present in real materials – could in fact help stabilizing a QSI ground state in $Pr_2Hf_2O_7$.

Having all these elements in mind, it makes sense to discuss here the origin of the continuum of excitations measured in $Pr_2Hf_2O_7$. Magnetic monopoles cannot usually be excited by scattering neutrons from a non-Kramers ion such as $Pr^{3+}$ [33]. However, magnetic monopoles could be introduced where some sort of residual disorder mixes the dipolar and quadrupolar components of the ground state. The continuum observed in our experiments could also originate in the dual, electric charges of the gauge theory, as was suggested in a recent theory work [33]. Finally, one should note here as well that there are other possible explanations regarding the origin of the observed continuum, which we have considered but could be ruled-out – as stated in more details in our original paper [18].

## COMPARISON WITH OTHER PYROCHLORE MATERIALS

The existence of "quantum effects" in pyrochlore materials is not limited to the theoretical understanding and experimental search for the QSI state. Other examples include materials such as, for instance, $Er_2Ti_2O_7$ – where quantum fluctuations may play a role in the 'order by disorder' mechanism [34-36]; $Yb_2Ti_2O_7$ – where an exotic long-range order retains dynamics down to the lowest temperatures [28,36-38]; or $Nd_2Zr_2O_7$ – where a peculiar ratio of exchange interactions between different pseudo-spin components leads to the fragmentation of the degrees of freedom into a fluctuating Coulomb phase and a long-range ordered phase of crystallized monopoles [15-16,39].

Different pyrochlore materials based on non-Kramers rare-earths were extensively studied with the aim to find evidence for a QSI ground state. The most studied cases are $Tb_2Ti_2O_7$ and $Pr_2Zr_2O_7$.

$Tb_2Ti_2O_7$ has a rich physics [40-43] where the lattice strongly couples to a spin system that lacks the magnetic long-range order predicted at around 1 K for this Ising antiferromagnet. Recent experiments suggest that electric quadrupole-ordered phases compete with spin liquid ground states depending on the exact stoichiometry of the sample [44-46].

Tb$_2$Ti$_2$O$_7$ remains an interesting candidate for the QSI state, whose origin might be related to quantum-mechanical processes allowed by a low-lying crystal-electric field level found around 1 meV [47].

Pr$_2$Zr$_2$O$_7$ is the pyrochlore material that is of course most closely related to our study. Tetravalent Zr and Hf cations are indeed quite close in terms of electronegativities and ionic radii, so that the corresponding Pr-based pyrochlore phases might be expected to have the same properties. However, hafnium oxides are even more refractory materials than zirconium oxides, which influences the optimal conditions of the traveling-solvent floating-zone (TSFZ) crystal growth. Adding the fact that crystals were grown at various institutions using different equipment, the result is that single-crystal samples of Pr$_2$Zr$_2$O$_7$ and Pr$_2$Hf$_2$O$_7$ have proved different in their level of structural disorder. Given the potential influence of disorder around non-Kramers ions [30], a short summary can be that the results of previous studies on Pr$_2$Zr$_2$O$_7$ differ from our work on Pr$_2$Hf$_2$O$_7$ in the extent to which samples reflect disorder in the physics of the system. In particular, it seems likely that strong disorder in samples of Pr$_2$Zr$_2$O$_7$ leads to degrees of freedom having essentially a quadrupolar character [48-50] (although other scenarios have been proposed [51]), while the reduced level of disorder in Pr$_2$Hf$_2$O$_7$ preserves magnetic dipoles and introduces quantum fluctuations on the spin ice manifold [18]. A clear experimental signature of this difference is that spin ice-like scattering is found in the quasi-elastic channels in Pr$_2$Hf$_2$O$_7$ [18] – as expected from a QSI, while all studies on Pr$_2$Zr$_2$O$_7$ have essentially shown signals centered on finite energy transfers around 0.3 meV [48-49,51-53].

It should be noted here that the quasi-elastic map published in the first study of Pr$_2$Zr$_2$O$_7$, by Kimura et al. [52], interpreted this signal as evidence for short-range spin ice correlations static on the scale of 2 picoseconds – corresponding to the finite energy-resolution of their experiment. Such a resolution (equivalently 0.38 meV) thus integrates over a range including the energy scale of the interactions in Pr-based pyrochlores, where inelastic scattering can be expected. This isn't an issue for the Pr$_2$Zr$_2$O$_7$ data as it was found that there is essentially no quasi-elastic contribution in the studied samples [48-49,51-53]. However, it is worth making this point when comparing with Pr$_2$Hf$_2$O$_7$, where the resolution of our measurements on IN5 (0.050 meV) is crucial for a clean separation of quasi-elastic and inelastic scattering [18]. It turns out that a quasi-elastic signal of the type expected in a QSI is measurable in Pr$_2$Hf$_2$O$_7$, and that its q-dependence is distinguishable from the "starfish" scattering observed at finite energy transfers and ascribed to the gapped excitations of the QSI state.

Finally, we should certainly mention here the recent interest on Ce$^{3+}$-based pyrochlores. On the basis of bulk measurements and muon spin relaxation experiments, Ce$_2$Sn$_2$O$_7$ was identified as a candidate material for the realization of a quantum spin liquid state on the pyrochlore lattice [54]. It was then pointed by theoreticians that the bulk properties support the existence of a 'dipole-octupole' ground state doublet [55], meaning that the degrees of freedom are magnetic octupoles in directions perpendicular to the local trigonal direction of the magnetic dipoles. Recently, neutron scattering experiments were reported on the related zirconate compound [56-57]. Though being at an early stage, these results further confirm the interest of $Ce^{3+}$-based pyrochlores.

## CONCLUSIONS AND OUTLOOK

Our neutron scattering study of $Pr_2Hf_2O_7$ reveals signatures of a quantum spin ice ground state. Such observations constitute a concrete example of a three-dimensional quantum spin liquid – a topical state of matter that has so far mostly been observed in lower dimensions. First, the mapping of the quasi-elastic structure factor at 0.05 K in this material reveals pinch points (Figures 2 and 3) – a signature of a classical spin ice – that are partially suppressed, as expected in a quantum spin ice. The line shape of the pinch-point scattering was compared with calculations of a lattice field theory of a quantum spin ice, in which low-energy gapless photon excitations explain the broadening of the curve. This result allows an estimate for the speed of light associated with magnetic photon excitations. Second, our data also reveal a continuum of inelastic spin excitations (Figure 4), which resemble predictions for the fractionalized, topological excitations of a quantum spin ice. Taken together, these two signatures strongly suggest that the low-energy physics of $Pr_2Hf_2O_7$ can be described by emergent quantum electrodynamics.

Further experimental work is needed to fully characterize the low temperature state of $Pr_2Hf_2O_7$. This includes careful measurements of the heat capacity down to very low temperature in order to compare its temperature dependence with predictions for linearly-dispersing photons, and to determine the entropy associated with the spin-liquid state. Moreover, detailed predictions of how pinch points evolve with temperature exist and such measurements would provide important information about quantum coherence in the QSI state. Finally, directly probing the photons through higher resolution techniques, using for example neutron spin echo or back-scattering instruments, is one of the ultimate experimental goals in a QSI material.

## REFERENCES

[1] Pauling, L. *J. Am. Chem*. Soc. **57**, 2680 (1935).
[2] Castelnovo, C., Moessner, R. & Sondhi, S. L. *Annu. Rev. Condens. Matter Phys*. **3**, 35–55 (2012).
[3] Balents, L. Nature **464**, 199–208 (2010).
[4] Savary, L & Balents, L. *Rep. Prog. Phys.* **80**, 016502 (2016).
[5] Anderson, P. W. *Mat. Res. Bull.* **8**, 153 (1973).
[6] Tennant, D. A., Perring, T. G., Cowley, R. A., & Nagler, S. E. *Phys. Rev. Lett,* **70** 4003–4006 (1993).
[7] Han, T.-H. *et al. Nature* **492**, 406–410 (2012).
[8] Banerjee, A. *et al. Nature Mater*. **15**, 733–740 (2016).
[9] Shen, Y. *et al. Nature* **540**, 559–562 (2016).
[10] Gingras, M. J. P. and McClarty, P. A., *Rep. Prog. Phys.* **77**, 056501 (2014).